\documentclass[aps,prl,showpacs,twocolumn,amsmath,amssymb]{revtex4}
\usepackage[english]{babel}
\usepackage{latexsym}
\usepackage{graphics}
\usepackage{subfigure}
\usepackage{epsfig}

\begin{document}

\title{Direct Observation of the Class-B to Class-A Transition in the Dynamical Behavior of a Semiconductor Laser}

\author{Ghaya Baili,$^1$ Mehdi Alouini,$^{1,2}$ Thierry Malherbe,$^1$ Daniel Dolfi,$^1$ Isabelle Sagnes,$^{3}$ and Fabien Bretenaker$^{4}$}
\email{Fabien.Bretenaker@lac.u-psud.fr}

\affiliation{ $^1$Thales Research \& Technology France, RD 128, 91767 Palaiseau Cedex, France
\\
$^2$Institut de Physique de Rennes, CNRS-Universit\'e de Rennes I, 35042 Rennes Cedex, France
\\
$^3$Laboratoire de Photonique et de Nanostructures, CNRS, 91460 Marcoussis, France
\\
$^4$Laboratoire Aim\'e Cotton, CNRS-Universit\'e Paris Sud 11, 91405 Orsay Cedex, France}

\date{\today}


\begin{abstract}
The transition between the class-B and class-A dynamical behaviors of a semiconductor laser is directly observed by continuously controlling the lifetime of
the photons in a cavity of sub-millimetric to centimetric length. It is experimentally and theoretically proved that the transition from a resonant to an
overdamped behavior occurs progressively, without any discontinuity. In particular, the intermediate regime is found to exhibit features typical from both the
class-A and class-B regimes. The laser intensity noise is proved to be a powerful probe of the laser dynamical behavior.
\end{abstract}

\maketitle Since their discovery, lasers have been considered to be among the most exciting dynamical systems according to the wide variety of behaviors they
offer. Laser dynamics is so rich that it became a tool of choice to analyze other dynamical systems even in new area of physics. For instance, a fruitful
analogy can be found between Bose-Einstein condensation and laser phase transition \cite{Scully1999}. Although the laser is considered as a system far away
from thermal equilibrium, tremendous theoretical and experimental studies have been carried out in order to find a thermodynamic reinterpretation of most laser
phenomena. For instance, when the electromagnetic field is taken as the order parameter and the population inversion plays the role of temperature, laser
threshold appears as a second-order phase transition \cite{Graham1970,DeGiorgio1970}. Similar analogy is found in an active micro-cavity whose dimension is
inferior to wavelength. It is shown that the transition occurring from spontaneous emission enhancement/inhibition, due to confinement, into collective
stimulated emission can be reinterpreted as second order phase transition by analogy with ferromagnetism and superconductivity \cite{DeMartini1988}. First
order transition can also be observed in lasers. For instance, homogenously broadened lasers can sustain the oscillation of two bistable optical fields. In
this case, the laser field switching behaves as a first order transition \cite{Lett1981}.

More generally, it has been well established for long that the dynamical behavior complexity of a system increases as the number of degrees of freedom
increases, leading even to chaotic dynamics \cite{Berge1987}. In the peculiar case of lasers, chaos can be obtained provided that more than two degrees of
freedom are present in the system. Practical examples include lasers on which an external optical field, gain or loss feedback is applied in order to increase
the number of degrees of freedom \cite{Arecchi1986,Pieroux1994} or some molecular far infrared laser. In the case of single mode lasers, the number of degrees
of freedom is determined by the time scale of the system constants, namely (i) the active medium polarization decay rate $\gamma_{\bot}$, (ii) the population
decay rate $\gamma_{\parallel}$, and (iii) the cavity decay rate $\gamma_{\mathrm{cav}}$. Indeed, within the semi-classical approximation in which the atoms
are treated quantum mechanically but the optical field is treated classically, the Maxwell-Bloch equations leads to five differential nonlinear equations whose
resolution difficulty depends upon the time scale of $\tau_{\bot}=1/\gamma_{\bot}$, $\tau_{\parallel}=1/\gamma_{\parallel}$ and
$\tau_{\mathrm{cav}}=1/\gamma_{\mathrm{cav}}$. Following the early classification of Arecchi et al. \cite{Arecchi1984}, class-C lasers are those for which the
three decay rates are of the same order of magnitude. This class includes some molecular far infrared lasers. Solving the Maxwell-Bloch equations gives rise to
a large number of solutions including chaotic behaviors. In class-B lasers the active medium polarization decays so rapidly that it can be adiabatically
eliminated from the Maxwell-Bloch equations. The number of degrees of freedom is reduced to two, leading to a couple of so-called rate equations. Class-B
includes most of the lasers used today such as solid-state lasers and semiconductor lasers. Finally, when the active medium polarization and population
inversion both decay much faster than the optical field, the population inversion can be eliminated adiabatically as well. The system has one degree of freedom
and the laser dynamics is then ruled by a single field equation. Most atomic gas lasers belong to this family.

The purpose of the present paper is to explore experimentally how the transition from class-B to class-A occurs. To this aim, we intend to
probe the laser dynamics while the system evolves from two degrees of freedom to one degree of freedom. Although this transition can be
modeled without too much difficulty, it has been not yet observed experimentally. Thus the agreement between the theoretical predictions and
experimental results is still an open question.
\begin{figure}
\begin{center}
\includegraphics[width=8.6 cm]{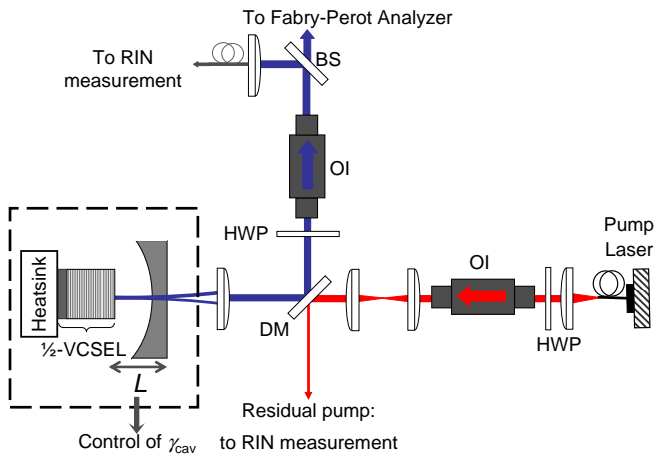}
\end{center}
\caption{Sketch of the experiment. The control of the cavity length $L$ permits to control the photon decay rate $\gamma_{\mathrm{cav}}$. OI's are optical
isolators, DM is a dichroic mirror, BS is a beam-splitter, and HWP's are half-wave plates.}\label{Figure 01}
\end{figure}
Direct observation of class-B to class-A transition is not easy to handle experimentally because it relies on finding technical solutions to
a certain number of constraints. As a starting point, the population inversion and cavity decay rates must be of the same order of
magnitude. This situation can be reached using a semiconductor active medium in conjunction with a few millimeter-long high-finesse optical
cavity. Second, one must be able to tune continuously one of the two decay rates while keeping the other parameters of the laser constant.
The obvious approach is to adjust the photon cavity lifetime rather than the population inversion lifetime. Keeping in mind that the laser
parameters must remain constant during class-A to class-B transition, the only way to change the cavity lifetime without modifying the other
laser parameters, such as threshold and pumping rate, is to adjust the cavity length. Finally, the laser must remain single frequency
within, and in between, the two boundary situations namely, class-A oscillation (long cavity) and class-B oscillation (short cavity). All
these constraints can be fulfilled using a semiconductor active medium inserted into a dedicated tunable external cavity, as sketched in
Fig.\;\ref{Figure 01}. Following this approach, the active medium we have chosen is a half-Vertical Cavity Surface Emitting Laser
(half-VCSEL). A design based on a surface-emitting semiconductor is preferred to that of edge-emitting semiconductor because pure single
mode operation is easier to obtain, in particular when the laser cavity becomes long. Furthermore, it is worthwhile to notice that common
Vertical Cavity Surface Emitting Lasers (VCSEL) belong to the class-B family. Their cavity length being in the micrometer range, the photon
lifetime is shorter than the population inversion lifetime. Thus, their dynamics behave as a second order filter exhibiting damped
relaxation oscillations \cite{Halbritter2004}. On the other hand, our recent experiments on intensity noise reduction in external cavity
VCSELs have confirmed that when the optical cavity is long enough, so that the photon lifetime gets longer than the carrier lifetime, the
laser dynamics behave as a low-pass first-order filter \cite{Baili2008} proving that the laser operates in the class-A regime. Given that
the two lasers in refs. \cite{Halbritter2004} and \cite{Baili2008,Baili2007} have active media of same nature whereas they exhibit two
different dynamics, using a half-VCSEL in order to achieve a continuous transition from class-A to class-B regime is a good starting point.

In our experiment (see Fig.\;\ref{Figure 01}), we have thus used the half-VCSEL whose structure is described in refs. \cite{Baili2007,Baili2008}. When inserted
inside an optical cavity and pumped at 808~nm, this structure oscillates at 1000~nm. To be able to reach sub-mm cavity lengths, the quantum wells in the active
medium are pumped through the laser output coupler, which has a radius of curvature of 25~mm and a transmission equal to 1\% at 1000~nm. In these conditions,
the laser beam diameter for a 1-mm cavity length is of the order of 100~$\mu$m. The pump beam, provided by a 3~W fiber-coupled diode, is focused to the same
diameter on the structure using a set of four lenses, as shown in Fig.\;\ref{Figure 01}, forcing the laser to oscillate in a single TEM$_{00}$ transverse mode.
The laser beam at the output of the VECSEL is isolated using a dichroic mirror DM and then analyzed. We check that the VECSEL oscillates in a single
longitudinal mode.
\begin{figure}
\begin{center}
\includegraphics[width=8.6 cm]{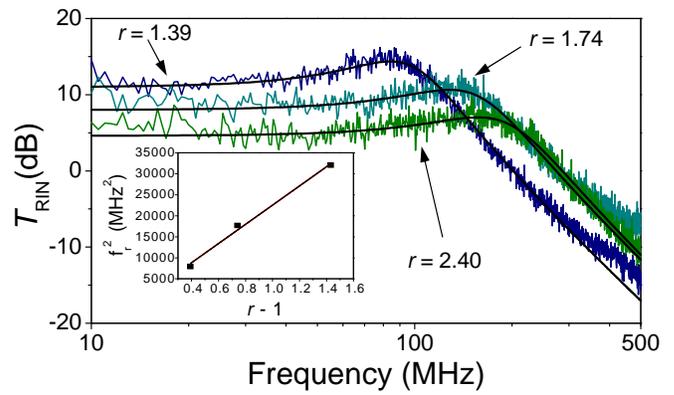}
\end{center}
\caption{Noise transfer function versus frequency for a cavity length $L=0.85\;\mathrm{mm}$ for three values of the relative pumping rate $r$. The noisy curves
are measurements. The smooth ones are fits obtained using Eq.\;\ref{equation01}. Insert: evolution of the square of the relaxation oscillation frequency versus
$r-1$.}\label{Figure 02}
\end{figure}

\begin{figure*}
\begin{center}
\includegraphics[width=17.8 cm]{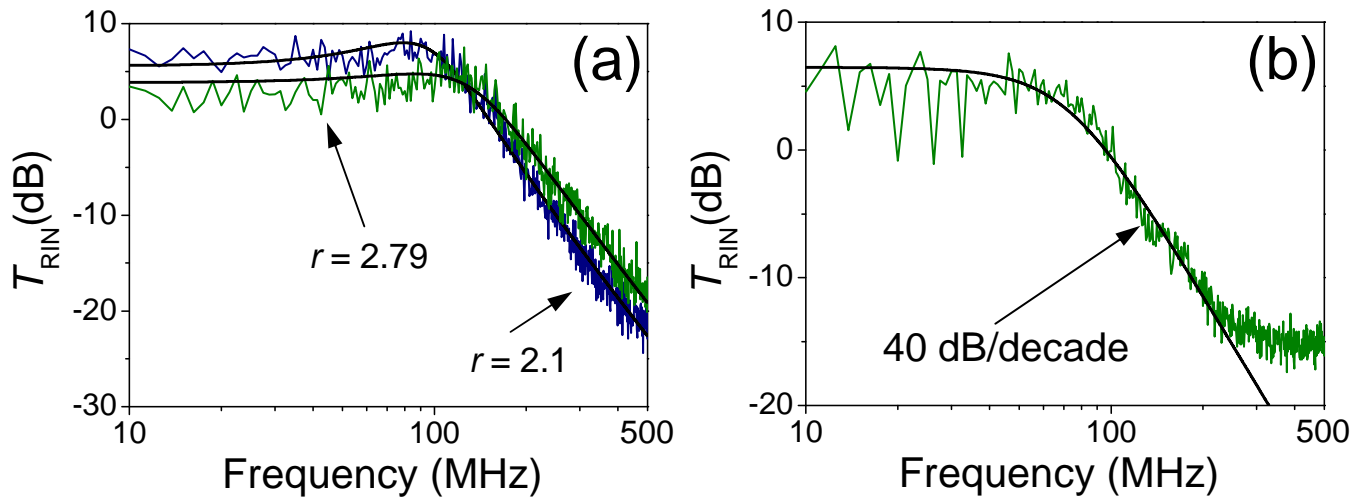}
\end{center}
\caption{(a) Same as Fig.\;\ref{Figure 02} for a cavity length $L=1.26\;\mathrm{mm}$ and $r=2.1$ and $2.79$. (b)  Same as Fig.\;\ref{Figure 02} for
$L=2.0\;\mathrm{mm}$ and $r=1.9$. The plateau observed from 300 to 500~MHz is due to the measurement noise floor.}\label{Figure 03}
\end{figure*}

We first choose a cavity short enough for the laser to exhibit a class-B dynamics. To this aim, we adjust the cavity length down to $L=0.85\,\mathrm{mm}$. In
these conditions, we expect the cavity photon lifetime $\tau_{\mathrm{cav}}$ to lie between 0.3 and 0.6~ns. Indeed, the cavity round-trip losses must be
between 1\% (transmission of the output coupler) and 2\% (maximum gain of our half-VECSEL). Since the carrier lifetime $\tau_{\parallel}$ is of the order of
3~ns, we expect our VECSEL to behave like a class-B laser with a relaxation oscillation frequency $f_{\mathrm{r}}$ of the order of 100~MHz. To monitor the
dynamical behavior of the laser, rather than measuring its modulation transfer function by modulating its gain or losses, we deduce this transfer function by
observing how the pump laser intensity noise is transferred to the laser intensity noise \cite{Yu1987}. It is indeed well known that the intensity noise of the
laser is a good probe of the laser dynamics \cite{McCumber1966}, provided the measured noise is well above the shot noise limit. We thus measure the VECSEL
relative intensity noise (RIN) spectrum and divide it by the measured RIN spectrum of the pump laser (see Fig.\;\ref{Figure 01}). We call $T_{\mathrm{RIN}}$
the RIN transfer function, i.e., the ratio of these RIN values.  Typical measurements of $T_{\mathrm{RIN}}$ versus noise frequency are reproduced in
Fig.\;\ref{Figure 02} for three values of the relative excitation $r$ (pump power normalized to threshold) of the laser. These transfer functions exhibit the
typical shape expected from a second-order resonant filter with a 40~dB/decade roll-off. Such a behavior is a signature of the class-B regime. The experimental
spectra of Fig.\;\ref{Figure 02} are fitted using the following expression derived for class-B lasers \cite{Baili2008}:
\begin{equation}
T_{\mathrm{RIN}}(f)=\frac{\gamma_{\parallel}^2\gamma_{\mathrm{cav}}^2 r^2}{\left[\gamma_{\parallel}\gamma_{\mathrm{cav}}(r-1)-(2\pi
f)^2\right]^2+\left(2\pi f\gamma_{\parallel}r\right)^2}\ .\label{equation01}
\end{equation}
Using Eq.\;\ref{equation01}, the fits of Fig.\;\ref{Figure 02} lead to $\tau_{\parallel}=3.3\;\mathrm{ns}$ and $\tau_{\mathrm{cav}}=0.31\;\mathrm{ns}$ for
$r=1.39$, $\tau_{\parallel}=1.9\;\mathrm{ns}$ and $\tau_{\mathrm{cav}}=0.31\;\mathrm{ns}$ for $r=1.74$, and $\tau_{\parallel}=2.4\;\mathrm{ns}$ and
$\tau_{\mathrm{cav}}=0.40\;\mathrm{ns}$ for $r=2.4$. The corresponding values of the relaxation oscillation frequency $f_{\mathrm{r}}$ are 89, 133, and
179~MHz, respectively. The evolution of $f_{\mathrm{r}}$ versus $r$ is reproduced in the inset in Fig.\;\ref{Figure 02}. It confirms that $f_{\mathrm{r}}^2$
evolves linearly with $(r-1)$, as expected for a class-B laser. Moreover, the value of the photon lifetime $\tau_{\mathrm{cav}}$ deduced from the fits is
consistent with our initial guess and is ten times shorter than the carrier lifetime, proving also that the laser with $L=0.85\;\mathrm{mm}$ is clearly a
class-B laser. The variations in the values of $\tau_{\parallel}$ deduced from the fits can be attributed to measurement uncertainties and also to som
dependence of this effective lifetime on the pump power.

In order to get closer to the class-B to class-A transition, we increase the photon lifetime from about 0.3~ns to about 0.8~ns by increasing
the cavity length up to 1.26~mm. The corresponding transfer function is reproduced in Fig.\;\ref{Figure 03}(a) for two values of $r$. The
fits using Eq.\;\ref{equation01} lead to $\tau_{\parallel}=4.0\;\mathrm{ns}$ and $\tau_{\mathrm{cav}}=0.71\;\mathrm{ns}$ for $r=2.1$, and
$\tau_{\parallel}=3.0\;\mathrm{ns}$ and $\tau_{\mathrm{cav}}=0.82\;\mathrm{ns}$ for $r=2.79$. One can notice that the relaxation
oscillations are barely visible on the two spectra of Fig.\;\ref{Figure 03}(a), showing that we are closer to the class-A regime than in
Fig.\;\ref{Figure 02}.
\begin{figure}
\begin{center}
\includegraphics[width=8.6 cm]{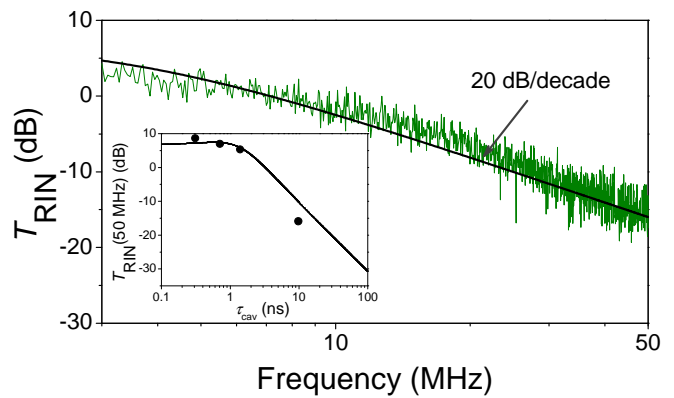}
\end{center}
\caption{Same as Fig.\;\ref{Figure 02} for $L=44\;\mathrm{mm}$ and $r=1.8$. The fit has been obtained using Eq. \ref{equation02}. Inset: theoretical (full
line) and experimental (dots) evolution of the RIN transfer function at $f=50\ \mathrm{MHz}$ versus photon lifetime.}\label{Figure 04}
\end{figure}

We go one step further by increasing the cavity length up to $L=2.0\;\mathrm{mm}$. The corresponding measured transfer function is reproduced in
Fig.\;\ref{Figure 03}(b). We can see that the resonance has disappeared (compare with Figs.\;\ref{Figure 03}(a) and \ref{Figure 02}). The transfer function now
looks like a low-pass filter, a feature usually considered to be typical of class-A laser. However, the roll-off is still equal to 40~dB/decade, which is
typical of class-B lasers \cite{Verdeyen1995}. We are thus exactly in the intermediate case in which the laser behavior exhibits features from both the class-A
and class-B regimes. This is consistent with the fact that the fit using Eq.\;\ref{equation01} gives values of $\tau_{\parallel}$ and $\tau_{\mathrm{cav}}$
which are of the same order of magnitude ($\tau_{\parallel}=2.8\;\mathrm{ns}$ and $\tau_{\mathrm{cav}}=1.4\;\mathrm{ns}$). It is worth noting that the
vanishing of the relaxation oscillations in the spectrum of Fig.\;\ref{Figure 03}(b) is really a signature of a modification of the laser dynamics at the
border between the class-A and class-B regimes. It is different from the overdamping of relaxation oscillations that occurs in diode lasers due to spontaneous
emission or to gain compression \cite{Petermann1991}.

When the laser becomes really a class-A laser, then $\gamma_{\mathrm{cav}}\ll\gamma_{\parallel}$, and, for $f\ll\gamma_{\parallel}/2\pi$,
Eq.\;\ref{equation01} becomes:
\begin{equation}
T_{\mathrm{RIN}}(f)=\frac{\gamma_{\mathrm{cav}}^2}{\left[\gamma_{\mathrm{cav}}(\frac{r-1}{r})\right]^2+(2\pi f)^2}\ .\label{equation02}
\end{equation}
This is the transfer function of a first-order low-pass filter, with a cut-off frequency given by $(\frac{r-1}{r})\frac{\gamma_{\mathrm{cav}}}{2\pi}$ and a
20~dB/decade roll-off. To reach this regime, we increase the cavity length up to $L=44\;\mathrm{mm}$. The output coupler now has a 50~mm radius of curvature
and has again a 1\% transmission at 1000~nm. In this cavity, the structure is no longer pumped through the output coupler but from the side of the cavity (for
details, see \cite{Baili2007,Baili2008}). A 150~$\mu$m \'etalon forces the laser to operate in single-frequency regime. Now the laser transfer function
exhibits a 20~dB/decade roll-off, and can be fitted using Eq.\;\ref{equation02}, as shown by the full line in Fig.\;\ref{Figure 04}. It leads to
$\tau_{\mathrm{cav}}=9.8\;\mathrm{ns}$ (corresponding to 1.5\% losses per round-trip) which is indeed much longer that $\tau_{\parallel}$. This proves
definitely that the laser has now reached a pure class-A behavior. The transition from the class-B to class-A regime is also clearly seen the inset of
Fig.\;\ref{Figure 04}, which displays the evolution of the noise transfer function at a fixed frequency (50 MHz) with the photon lifetime. In this inset, the
theoretical plot has been obtained using Eq.\;\ref{equation01} with $r=1.89$ and $\tau_{\parallel}=2.9\;\mathrm{ns}$ and the experimental dots correspond to
$r$ close to 2. The abrupt decrease of $T_{\mathrm{RIN}}$ versus $\tau_{\mathrm{cav}}$ is a clear signature of the transition.

In conclusion, the transition from class-B to class-A dynamics has been directly observed by continuously modifying the photon lifetime in a dedicated single
mode laser cavity. We have confirmed that this transition occurs progressively as expected theoretically. Furthermore, we have been able to isolate an
intermediate regime in which the laser exhibits simultaneously features typical of the two regimes. Indeed, the relaxation oscillations disappear as expected
from class-A lasers, while the transfer function roll-off is still characteristic of class-B lasers. These observations have been made technically possible by
obtaining single mode oscillation in a very low losses sub-mm optical resonator including an active medium whose population inversion decay time is of the same
order of magnitude of the resonator decay time and whose polarization can be eliminated adiabatically. The laser intensity noise in such a system is shown to
be a powerful probe of the laser dynamics.

The control of the exact conditions in which a laser switches from class-A to class-B dynamics is important for the control of the intensity noise in
MEMS-VCSELs. Indeed, one wants the cavity to be as short as possible to extend the laser tuning range while keeping the laser quiet in order to maximize the
signal-to-noise ratio when the laser is used to probe absorption \cite{Lackner2006}. Controlling the nature of the laser dynamics is also important in
fundamental studies aiming at understanding the role played by the enhancement of the spontaneous emission in the laser relaxation oscillations and noise
\cite{Bjork1994}.
%
%\begin{acknowledgments}
%The authors are happy to thank L. Morvan, T. Merlet, and J. Chazelas for their help and support.
%\end{acknowledgments}

\end{document}